\documentstyle[graphicx, 12pt]{article}
\begin{document}

\small
\hoffset=-1truecm
\voffset=-2truecm
\title{\bf The thermodynamic quantities of a black hole with an $f(R)$ global monopole}
\author{Jingyun Man \hspace {1cm} Hongbo Cheng\footnote {E-mail address: hbcheng@ecust.edu.cn}\\
Department of Physics, East China University of Science and
Technology,\\ Shanghai 200237, China\\
The Shanghai Key Laboratory of Astrophysics, Shanghai 200234,
China}

\date{}
\maketitle

\begin{abstract}
The thermodynamic quantities such as the local temperature, heat
capacity, off-shell free energy and the stability of a black hole
involving a global monopole within or outside the $f(R)$ gravity
are examined. We compare the two classes of results to show the
influence from the generalization of the general relativity. It is
found that the $f(R)$ theory will modify the thermodynamic
properties of black holes, but the shapes of curves for
thermodynamic quantities with respect to the horizon are similar
to the results within the frame of general relativity. In both
cases there will exist a small black hole which will decay and a
large stable black hole in the case that the temperatures are
higher than their own critical temperature.
\end{abstract}
\vspace{7cm} \hspace{1cm} PACS number(s): 04.70Bw, 14.80.Hv

\newpage

\noindent \textbf{I.\hspace{0.4cm}Introduction}

Recently more contributions were paid to the thermodynamics of
various kinds of black holes. More than thirty years ago,
Bekenstein pointed out that the entropy of a black hole is
proportional to its surface area [1-3]. Hawking also discussed the
particle creation around a black holes to show that the black hole
has a thermal radiation with the temperature subject to its
surface gravity [4]. The issues about phase transitions of black
holes in the frame of semiclassical gravity were listed in Ref.
[5]. Further the thermodynamic properties of modified
Schwarszchild black holes have been investigated [6, 7]. The
thermodynamic behaviours including phase transition in
Born-Infeld-anti-de Sitter black holes were explored by means of
various ways [8, 9]. The phase transition of the quantum-corrected
Schwarzschild black hole was discussed, which fosters the research
on the quantum-mechanical aspects of thermodynamic behaviours
[10].

In the process of the vacuum phase transition in the early
Universe the topological defects such as domain walls, cosmic
strings and monopoles were generated from the breakdown of local
or global gauge symmetries [11, 12]. Among these topological
defects, a global monopole as a spherical symmetric topological
defect occurred in the phase transition of a system composed by a
self-coupling triplet of scalar field whose original global O(3)
symmetry is spontaneously broken to U(1). It has been shown that
the metric outside a monopole has a deficit solid angle [13]. We
researched on the strong gravitational lensing for a massive
source with a global monopole and find that the deficit angle
associated with the monopole affect the lensing properties [14].
H. A. Buchdahl proposed a modified gravity theory named as $f(R)$
gravity to explain the accelerated expansion of the Universe
instead of adding unknown forms of dark energy or dark matter
[15-18]. The metric outside a gravitational object involving a
global monopole in the context of $f(R)$ gravity theory has been
studied [19]. Further the classical motion of a massive test
particle around the gravitational source with an $f(R)$ global
monopole is probed [20]. We also examine the gravitational lensing
for the same object in the strong field limit [21].

Here we plan to investigate the thermodynamics of a static and
spherically symmetric black hole swallowing a global monopole or
an $f(R)$ global monopole. It is significant to understand the
$f(R)$ theory in a new direction. We wish to find the influences
from the modified gravity on the thermal properties of black
holes. First of all we introduce a black hole containing a global
monopole in the context of $f(R)$ gravity theory and certainly the
black hole metric will recover to be one of the metrics of black
hole with a deficit solid angle. We show the dependence of the
horizon on the mass and parameters describing the monopole and the
modification from $f(R)$ theory. We exhibit the thermodynamic
characteristics due to the parameters of the black hole. The
thermodynamic stability will also be checked. We are going to
discuss the results in the end.

\vspace{0.8cm} \noindent \textbf{II.\hspace{0.4cm}The
thermodynamics of black holes involving an $f(R)$ global monopole}

We adopt the line element,

\begin{equation}
ds^{2}=A(r)dt^{2}-B(r)dr^{2}-r^{2}(d\theta^{2}+\sin^{2}\theta
d\varphi^{2})
\end{equation}

\noindent which is static and spherical. In the $f(R)$ gravity
theory, the action introduced is,

\begin{equation}
I=\frac{1}{2\kappa}\int d^{4}x\sqrt{-g}f(R)+I_{m}
\end{equation}

\noindent where $f(R)$ is an analytical function of Ricci scalar
$R$ and $\kappa=8\pi G$, $G$ is the Newton constant, $g$ is the
determinant of metric and $I_{m}$ is the action of matter fields
which can be denoted as,

\begin{equation}
I_{m}=\int d^{4}x\sqrt{-g}{\cal L}.
\end{equation}

\noindent Here the Lagrangian for global monopole is [13],

\begin{equation}
{\cal
L}=\frac{1}{2}(\partial_{\mu}\phi^{a})(\partial^{\mu}\phi^{a})
-\frac{1}{4}\lambda(\phi^{a}\phi^{a}-\eta^{2})^{2}
\end{equation}

\noindent where $\lambda$ and $\eta$ are parameters. The ansatz
for the triplet of field configuration showing a monopole is
$\phi^{a}=\eta h(r)\frac{x^{a}}{r}$ with $x^{a}x^{a}=r^{2}$, where
$a=1, 2, 3$. $h(r)$ is a dimensionless function to be determined
by the equation of motion. This model has a global O(3) symmetry,
which is spontaneously broken to U(1). The field equation reads,

\begin{equation}
F(R)R_{\mu\nu}-\frac{1}{2}f(R)g_{\mu\nu}-\nabla_{\mu}\nabla_{\nu}F(R)
+g_{\mu\nu}\Box F(R)=\kappa T_{\mu\nu}
\end{equation}

\noindent with $F(R)=\frac{df(R)}{dR}$ and $T_{\mu\nu}$ is the
minimally coupled energy-momentum tensor. The field equation (5)
was solved under the weak field approximation that assumes the
components of metric tensor like $A(r)=1+a(r)$ and $B(r)=1+b(r)$
with $|a(r)|$ and $|b(r)|$ being smaller than unity [19]. Here the
modified theory of gravity corresponds to a small correction on
the general relativity like $F(R(r))=1+\psi(r)$ with
$\psi(r)\ll1$. It is clear that $F(R)=1$ is equivalent to the
conventional general relativity. Further the modification can be
taken as the simplest analytical function of the radial coordinate
$\psi(r)=\psi_{0}r$. In this case the factor $\psi_{0}$ reflects
the deviation of standard general relativity. The external metric
of the black hole with a global monopole is found finally [19,
20],

\begin{equation}
A=B^{-1}=1-8\pi G\eta^{2}-\frac{2GM}{r}-\psi_{0}r
\end{equation}

\noindent where $M$ is the mass parameter. It should be pointed
out that the parameter $\eta$ is of the order $10^{16}GeV$ for a
typical grand unified theory, which means $8\pi G\eta^{2}\approx
10^{-5}$. If we choose $\psi_{0}=0$ excluding the modification
from $f(R)$ theory, the metric (6) will recover to be the result
by M. Barriola et.al. [13] like,

\begin{equation}
ds^{2}=(1-8\pi G\eta^{2}-\frac{2GM}{r})dt^{2}
-\frac{dr^{2}}{1-8\pi G\eta^{2}-\frac{2GM}{r}}
-r^{2}(d\theta^{2}+\sin^{2}\theta d\varphi^{2}).
\end{equation}

The function $A(r)$ is plotted in Figure 1 for comparison of two
metrics for a global monopole and $f(R)$ monopole black hole.
According to the behavior of $f(R)$ curve, it is clear that the
black hole has two horizons, one is the inner horizon which is
supposed to be the event horizon $r_{H}$, the other one is the
outer horizon. The event horizons of two metrics seem to be the
same from Figure 1, because the given parameters $8\pi G\eta^{2}$
is very small, but the analytic expression of $r_{H}$ are quite
distinct. Solving the equation $A(r_{H})=0$, the event horizon of
metric (6) is located at,

\begin{equation}
r_{H}=\frac{(1-8\pi G\eta^{2})-\sqrt{(1-8\pi G\eta^{2})^{2}
-8\psi_{0}GM}}{2\psi_{0}}.
\end{equation}

\noindent It also gives the relation between the mass parameter
$GM$ and the event horizon $r_{H}$,

\begin{equation}
GM=\frac{1}{2}r_{H}(1-8\pi G\eta^{2}-\psi_{0}r_{H})
\end{equation}

\noindent for $f(R)$ monopole metric. It is seen from Figure 2
that there is a maximun $GM_{0}=\frac{(1-8\pi
G\eta^{2})^{2}}{8\psi_{0}}$ at $r_{H0}=\frac{1-8\pi
G\eta^{2}}{2\psi_{0}}$ where the inner and outer horizon meets.
The Hawking temperature can be obtained,

\begin{eqnarray}
T_{H}^{M}=T_{H}^{M}(\eta, \psi_{0})\nonumber\hspace{1cm}\\
=\frac{1}{4\pi}[-g^{tt}g^{rr}g'_{tt}]|_{r=r_{H}}\nonumber\hspace{1mm}\\
=\frac{1}{4\pi}(\frac{1-8\pi G\eta^{2}}{r_{H}}-2\psi_{0})
\end{eqnarray}

\noindent where the prime stands for the derivative with respect
to the radial coordinate $r$. The local temperature is given by
[22],

\begin{eqnarray}
T_{loc}^{M}=\frac{T_{H}^{M}}{\sqrt{A(r)}}\nonumber\hspace{10cm}\\
=\frac{1}{4\pi}(\frac{1-8\pi G\eta^{2}}{r_{H}}-2\psi_{0})
\sqrt{\frac{r}{\psi_{0}r_{H}^{2}-(1-8\pi G\eta^{2})r_{H} +(1-8\pi
G\eta^{2})r-\psi_{0}r^{2}}}.
\end{eqnarray}

\noindent Having omitted the influence from the modified gravity,
we obtain the local temperature as,

\begin{equation}
T_{loc}=\frac{1}{4\pi r_{H}}\sqrt{\frac{(1-8\pi G\eta^{2})r}
{r-r_{H}}}.
\end{equation}

\noindent The local temperatures for a global monopole and an
$f(R)$ global monopole are shown respectively in the Figure 3.
There is a minimal local temperature $T_{c}$, which means if the
local temperature is below $T_{c}$, no black hole exists. When the
temperatures for the two kinds of black holes are high enough,
there will both exist two black holes, one is small and the other
is large. From the calculation of the extrema of local
temperature, $(\frac{\partial T_{loc}}{\partial r_{H}})_{r}=0$,
the minimal local temperature can be obtained as,

\begin{equation}
T_{c}^{M}=\frac{\sqrt{r}}{\pi}(\frac{\psi_{0}}{(1-8\pi
G\eta^{2})^{\frac{2}{3}} -(1-8\pi G\eta^{2}-2\psi_{0}
r)^{\frac{2}{3}}})^{\frac{3}{2}}.
\end{equation}

The modifying factor $\psi_{0}$ from $f(R)$ leads the minimum of
local temperature lower than the one for global monopole black
hole,

\begin{equation}
T_{c}=\frac{3\sqrt{3}}{8\pi r}\sqrt{1-8\pi G\eta^{2}}
\end{equation}
for $8\pi G\eta^{2}=10^{-5}$, $r=10$ and $\psi_{0}=0.02$, then
$T_{c}^{M}=0.01836$, $T_{c}=0.02067$. The Figure 4 shows that the
critical temperature is a decreasing function of the modifying
factor $\psi_{0}$ denoting the influence from $f(R)$ gravity here.

According to Bekenstein's opinion, the entropy is proportional to
the area of event horizon and denoted as [1-3],

\begin{eqnarray}
S=\frac{A_{0}}{4}\nonumber\\
=\pi r_{H}^{2}
\end{eqnarray}

\noindent leading $dS=2\pi r_{H}dr_{H}$.

From the first law of thermodynamics $dE_{loc}=T_{loc}dS$, the
thermodynamical local energy can be derived as,

\begin{eqnarray}
E_{loc}^{M}=E_{0}+\int_{S_{0}}^{S}T_{loc}^{M}dS\nonumber\hspace{7cm}\\
=r\sqrt{(1-8\pi G\eta^{2})-\psi_{0}r}
-\sqrt{r(r-r_{H})}\sqrt{(1-8\pi G\eta^{2})-\psi_{0}(r+r_{H})}
\end{eqnarray}

\noindent Here $S_{0}$ represents that $M=0$. For simplicity
$E_{0}=0$. The thermodynamic local energy in the case not
belonging to $f(R)$ theory becomes,

\begin{equation}
E_{loc}=\sqrt{(1-8\pi G\eta^{2})r}(\sqrt{r}-\sqrt{r-r_{H}}).
\end{equation}

For the purpose of checking the stability of the black holes, we
should discuss their heat capacity,

\begin{eqnarray}
C^{M}=(\frac{\partial E_{loc}^{M}}{\partial T_{loc}^{M}})_{r}\nonumber\hspace{9cm}\\
=2\pi(r-r_{H})[(1-8\pi G\eta^{2})-2\psi_{0}r_{H}] [(1-8\pi
G\eta^{2})-\psi_{0}(r+r_{H})]\nonumber\hspace{1cm}\\
\times[(1-8\pi G\eta^{2})(\frac{r^{2}}{r_{H}^{2}}-3)\psi_{0}
+(1-8\pi G\eta^{2})^{2}\frac{3r_{H}-2r}{2r_{H}^{2}}
+2\psi_{0}^{2}r_{H}]^{-1}.
\end{eqnarray}

\noindent Similarly we choose the modifying factor $\psi_{0}=0$,
the heat capacity of black holes leads,

\begin{eqnarray}
C=(\frac{\partial E_{loc}}{\partial T_{loc}})_{r}\nonumber\hspace{0.5cm}\\
=4\pi\frac{r_{H}^{2}(r-r_{H})}{3r_{H}-2r}.
\end{eqnarray}

\noindent According to Eq. (18) and Eq. (19), we compare the heat
capacities of these two kinds of black holes in the Figure 5. The
shapes of the curves of heat capacities are similar, but can be
recognized explicitly. The expression of heat capacity for a
Schwarzschild black hole with a global monopole is exactly the
same as the one for original Schwarzschild black hole. In Figure
5, the $f(R)$ curve shifts from the conventional one. In every
case the relatively smaller horizon leads the heat capacity to be
negative and the positive capacity is due to the larger horizon,
which meaning that the larger black holes are stable. It should be
emphasized that only huge black holes can survive for long time
within the frame of $f(R)$ theory. When $r_{H}>\frac{1-8\pi
G\eta^{2}-[(1-8\pi G\eta^{2})(2\psi_{0} r-(1-8\pi
G\eta^{2}))^{2}]^{3/2}}{2\psi_{0}}$, the $f(R)$ heat capacity is
positive and a large black hole is stable. A small black hole
appears unstable for $0<r_{H}<\frac{1-8\pi G\eta^{2}-[(1-8\pi
G\eta^{2})(2\psi_{0} r-(1-8\pi G\eta^{2}))^{2}]^{3/2}}{2\psi_{0}}$
since the heat capacity is negative. A large Schwarzchild black
hole with a monopole is stable when $r_{H}>\frac{2r}{3}$, while it
is unstable on $0<r_{H}<\frac{2r}{3}$.

In order to explore the phase transition among the black holes, we
should derive their off-shell free energy. The off-shell free
energy can be defined as,

\begin{equation}
F_{off}^{M}=E_{loc}^{M}-TS
\end{equation}

\noindent where $E_{loc}^{M}$ is thermodynamic local energy from
Eq. (16), $S$ is the entropy of the black hole from Eq. (15) and
$T$ is an arbitrary temperature. The off-shell free energy of
black holes containing a global monopole governed by $f(R)$
gravity theory can be calculated as,

\begin{equation}
F_{off}^{M}=r\sqrt{(1-8\pi G\eta^{2})-\psi_{0}r}
-\sqrt{r(r-r_{H})}\sqrt{(1-8\pi G\eta^{2})-\psi_{0}(r+r_{H})} -\pi
r_{H}^{2}T.
\end{equation}

\noindent If we are not going to consider the deviation from
$f(R)$ theory, the Eq. (21) will become the expression of the
off-shell free energy for the black holes with only a deficit
solid angle like,

\begin{equation}
F_{off}=\sqrt{(1-8\pi G\eta^{2})r}(\sqrt{r}-\sqrt{r-r_{H}}) -\pi
r_{H}^{2}T.
\end{equation}

\noindent In Figure 6 the behaviour of the off-shell free energy
of black hole involving a global monopole proposed by Barriola and
Vilenkin [13] is shown as a function of the horizon under several
temperatures. Having considered extrema of off-shell free energy,
$(\frac{\partial F_{off}}{\partial r_{H}})_{r}=0$ or
$(\frac{\partial F^{M}_{off}}{\partial r_{H}})_{r}=0$, we give out
the critical temperatures which has already been mentioned above.
When the temperature is lower than the critical value, no black
hole will appear. For the temperature above the critical one the
large black holes are stable but the smaller ones are unstable.
The dependence of the free energy of black holes with an $f(R)$
global monopole on the horizon is plotted in the Figure 7 when the
temperature is chosen to be several values around their own
critical temperature. The shapes of the off-shell free energy are
similar to those in the Figure 6. When the temperature is lower
than critical value, there is no real root from the equation
$(\frac{\partial F_{off}}{\partial r_{H}})_{r}=0$ or
$(\frac{\partial F^{M}_{off}}{\partial r_{H}})_{r}=0$, therefore
no black hole will appear. There are two equal roots
$r_{H1}=r_{H2}=\frac{1-8\pi G\eta^{2}-[(1-8\pi
G\eta^{2})(2\psi_{0} r-(1-8\pi G\eta^{2}))^{2}]^{3/2}}{2\psi_{0}}$
at critical temperature. When $T>T_{c}$, two physically meaningful
roots exist. An unstable small black hole appears at event horizon
$r_{H}=r_{H1}$, and a stable large black hole appears at
$r_{H}=r_{H2}$.

\vspace{0.8cm} \noindent \textbf{III.\hspace{0.4cm}Discussion}

We search for the thermodynamic quantities of the black holes with
a global monopole in the context of $f(R)$ gravity theory. We also
obtain the thermodynamic quantities excluding the modifications
from $f(R)$ theory. We compare the two classes of results to show
how the generalized gravity corrects the original thermodynamic
quantities such as local temperature, heat capacity, off-shell
free energy. It should be pointed out that the
$f(R)$-modifications on these quantities are manifest. It is
interesting that the shapes of these quantities of black holes
controlled by two different gravity theories are similar although
the quantities' expressions are more different from each other. In
both cases the small black holes are unstable and the large ones
are stable when the temperature is higher than the critical
temperature and no black hole can exist as the temperature is
sufficiently low. The critical temperature has something to do
with the modified factor from $f(R)$ theory. Here we open a new
window to explore the $f(R)$ theory revising the Einstein Gravity.

In this paper, in what we pay more attention to is the
thermodynamic behaviours of a black hole with an $f(R)$ global
monopole, as the stablity of the black hole is dependent on them.
There are some methods proposed for probing a gravitational source
which has a solid deflict angle subject to $f(R)$ global monopole.
But one necessary condition is that the black hole has to be
stable enough to be observed. Through calculation of the heat
capacity, only when the event horizon is larger than $\frac{1-8\pi
G\eta^{2}-[(1-8\pi G\eta^{2})(2\psi_{0} r-(1-8\pi
G\eta^{2}))^{2}]^{3/2}}{2\psi_{0}}$, the black hole of an $f(R)$
global monopole can be stable. It is the condition for a large and
more stable black hole given by extrema of off-shell free energy
as well. Moreover, a stable black hole also requires the local
temperature higher than the critical temperature $T_{c}$.

\vspace{1cm}
\noindent \textbf{Acknowledge}

This work is supported by NSFC No. 10875043 and is partly
supported by the Shanghai Research Foundation No. 07dz22020.

\newpage
\begin{figure}
\setlength{\belowcaptionskip}{10pt} \centering
\includegraphics[width=15cm]{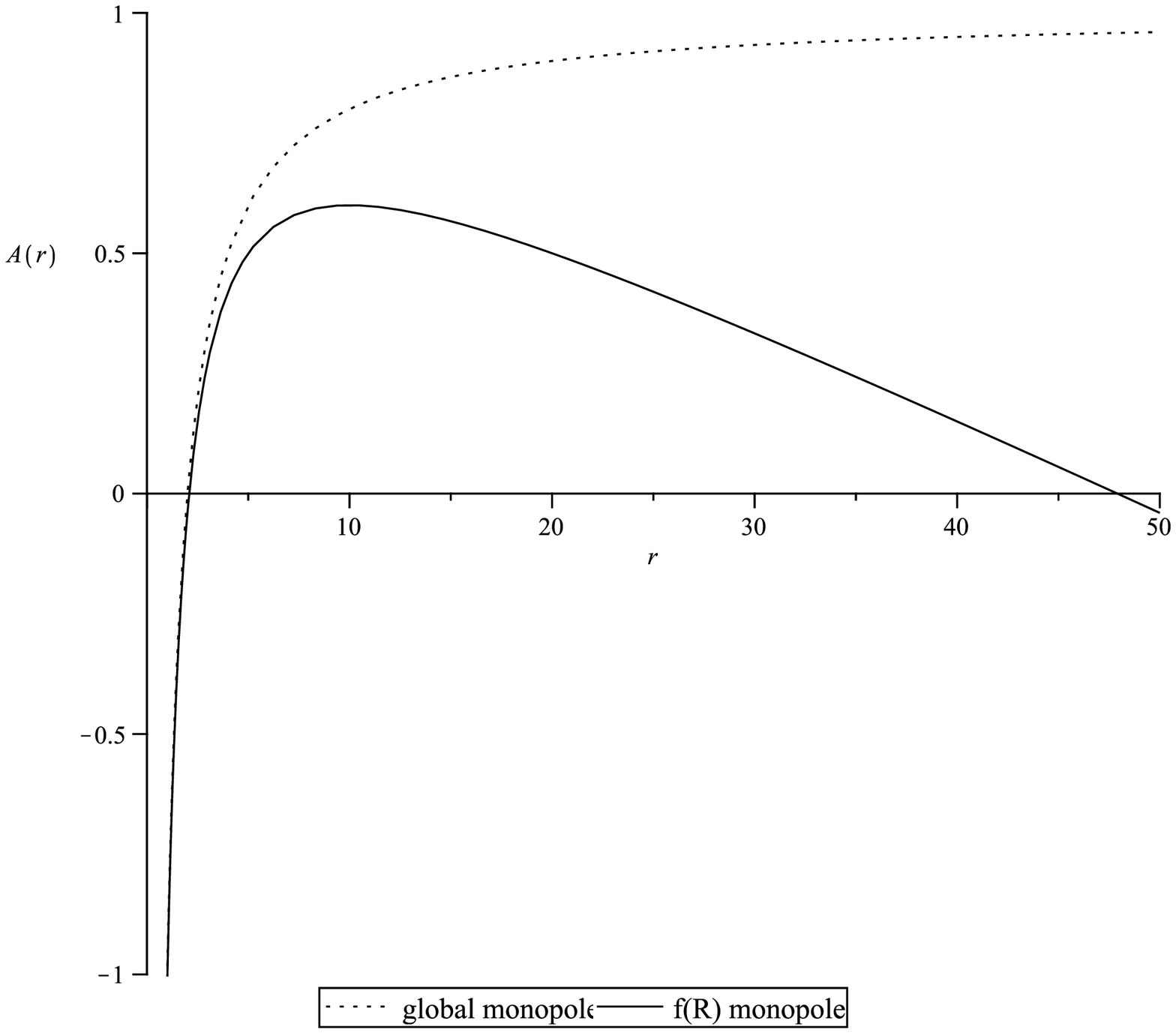}
\caption{The figure shows the function $A(r)$ of $f(R)$ monopole
black hole (solid line) and Schwarzchild black hole with a global
monopole (dot line). Here $8\pi G\eta^{2}\approx10^{-5}$,
$\psi_{0}=0.02$ and $GM=1$.}
\end{figure}

\newpage
\begin{figure}
\setlength{\belowcaptionskip}{10pt} \centering
  \includegraphics[width=15cm]{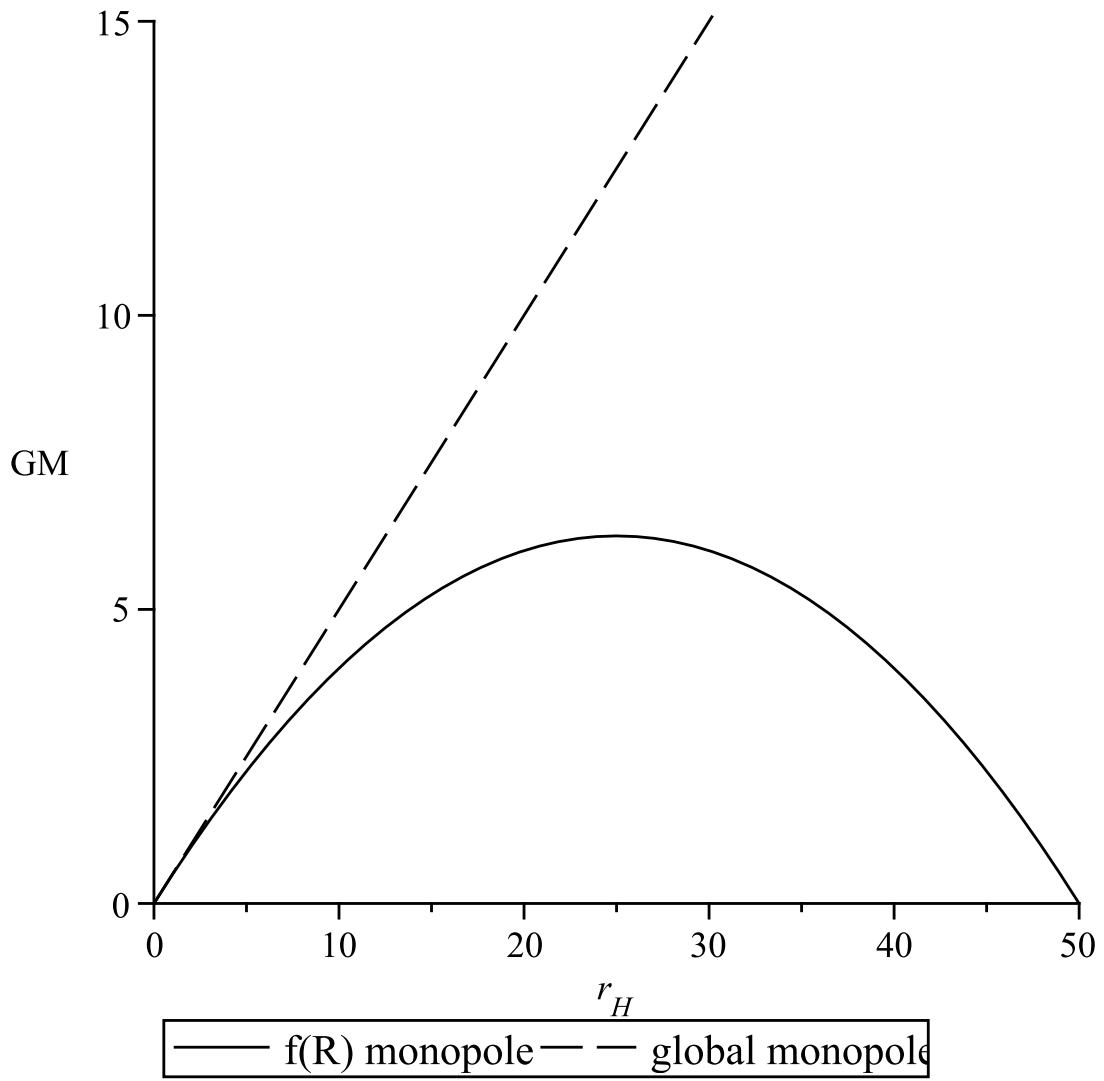}
  \caption{The figure show the mass parameter $GM$ for $f(R)$ monopole black hole (solid line) and global monopole black hole (dash line).
  Here $8\pi G\eta^{2}\approx10^{-5}$, $\psi_{0}=0.02$.}
\end{figure}

\newpage
\begin{figure}
\setlength{\belowcaptionskip}{10pt} \centering
  \includegraphics[width=15cm]{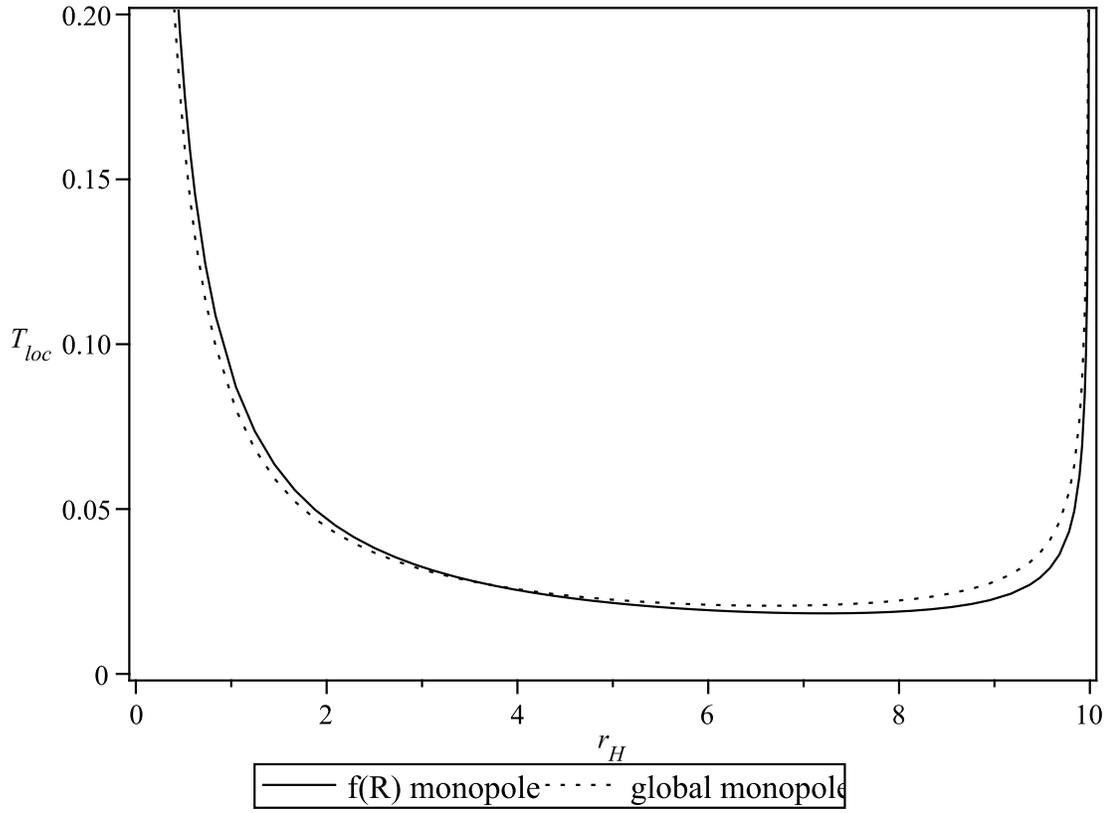}
  \caption{The solid and dotted curves of the dependence of the local temperatures
  on the horizon with $8\pi G\eta^{2}\approx10^{-5}$, $r=10$ and $\psi_{0}=0.02$
  for the Schwarzschild black hole with an $f(R)$ global monopole or a global monopole
respectively.}
\end{figure}

\newpage
\begin{figure}
\setlength{\belowcaptionskip}{10pt} \centering
  \includegraphics[width=15cm]{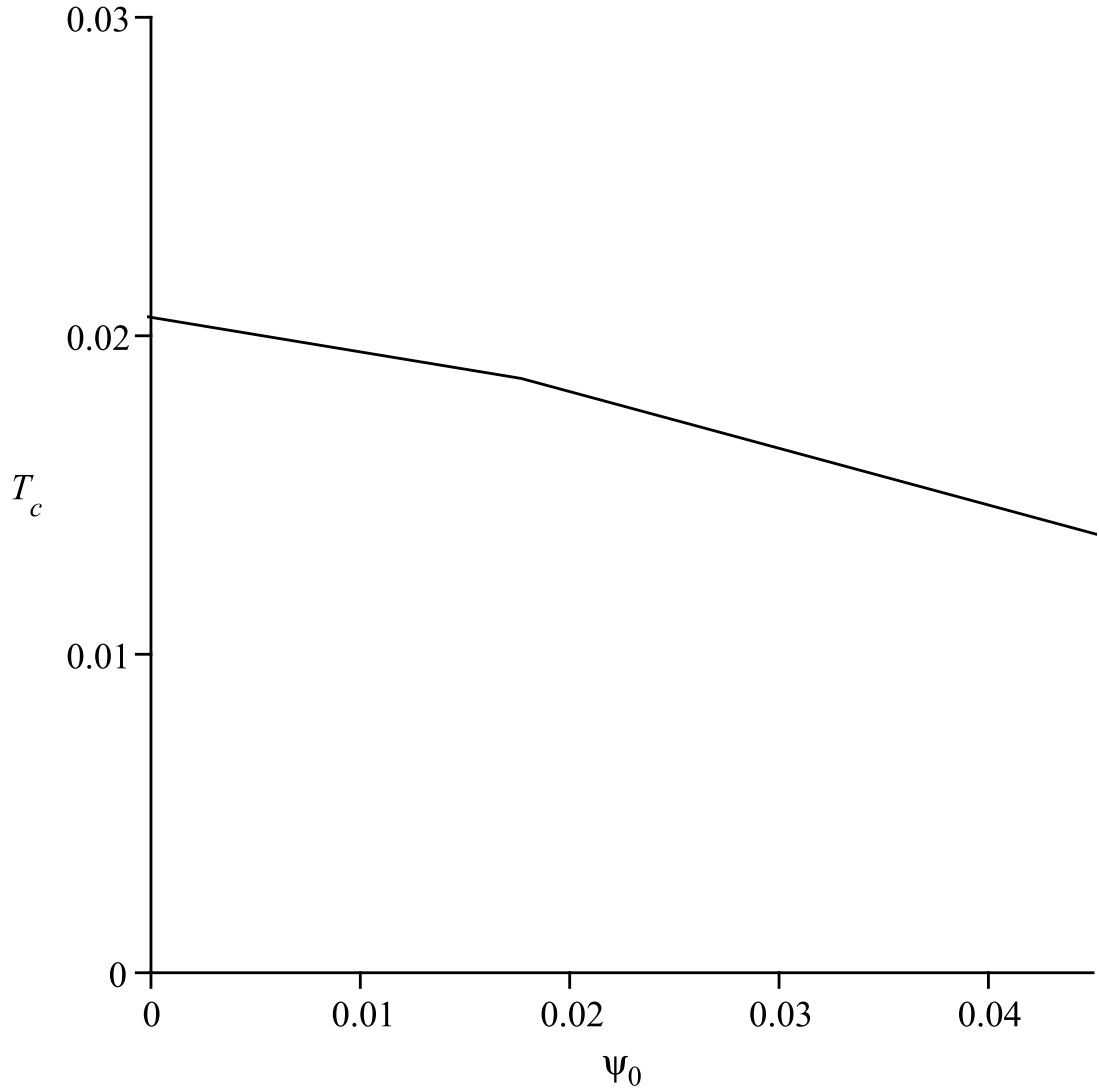}
  \caption{The dependence of the critical temperature for $8\pi G\eta^{2}\approx10^{-5}$
  and $r=10$ on the modifying factor $\psi_{0}$ from $f(R)$ gravity.}
\end{figure}

\newpage
\begin{figure}
\setlength{\belowcaptionskip}{10pt} \centering
  \includegraphics[width=15cm]{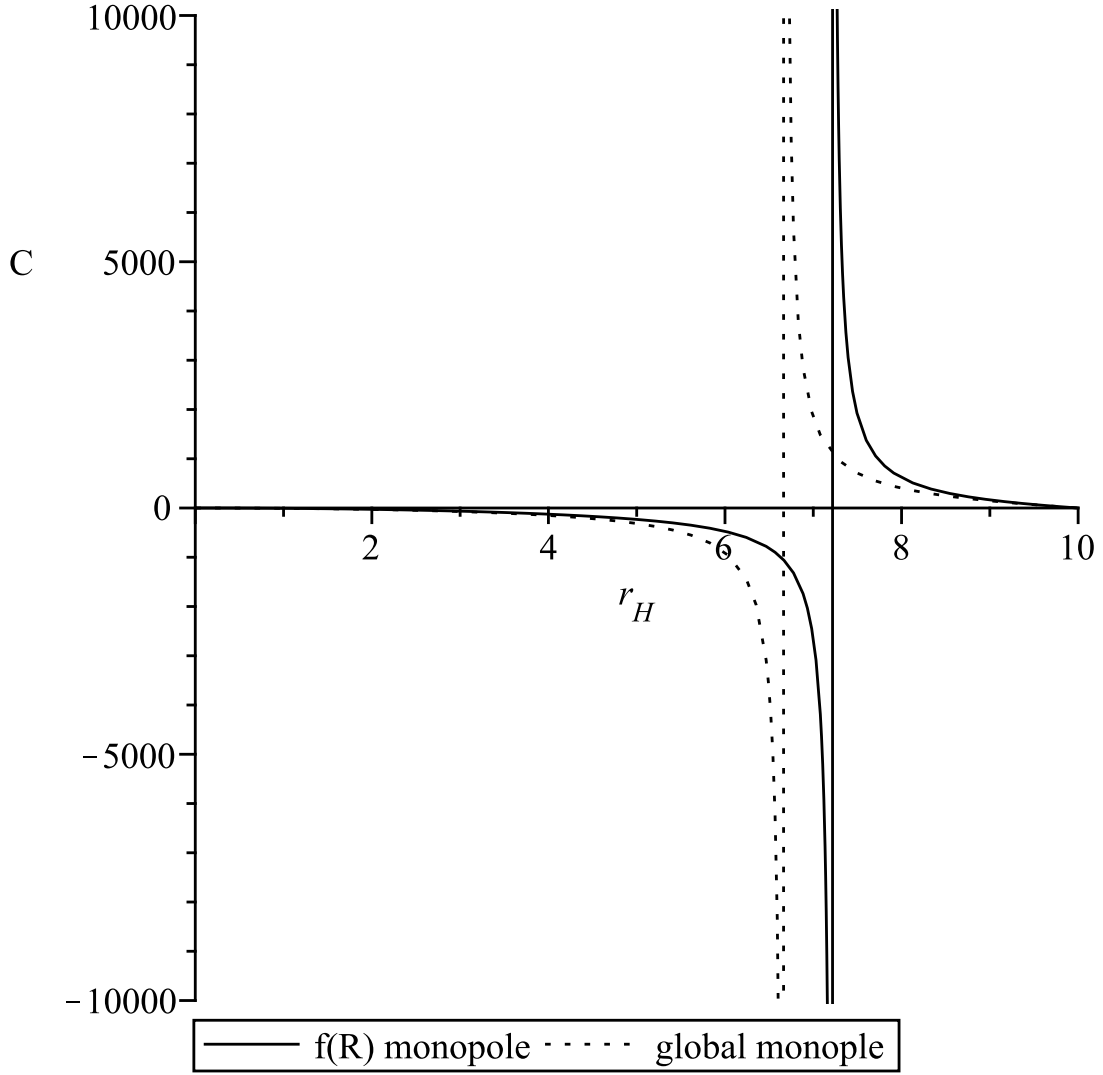}
  \caption{The solid and dotted curves correspond to the dependence of
  the heat capacities on the horizon with $8\pi G\eta^{2}\approx10^{-5}$, $r=10$ and $\psi_{0}=0.02$
  for the Schwarzschild black hole with an $f(R)$ global monopole or a global monopole
respectively.}
\end{figure}

\newpage
\begin{figure}
\setlength{\belowcaptionskip}{10pt} \centering
  \includegraphics[width=15cm]{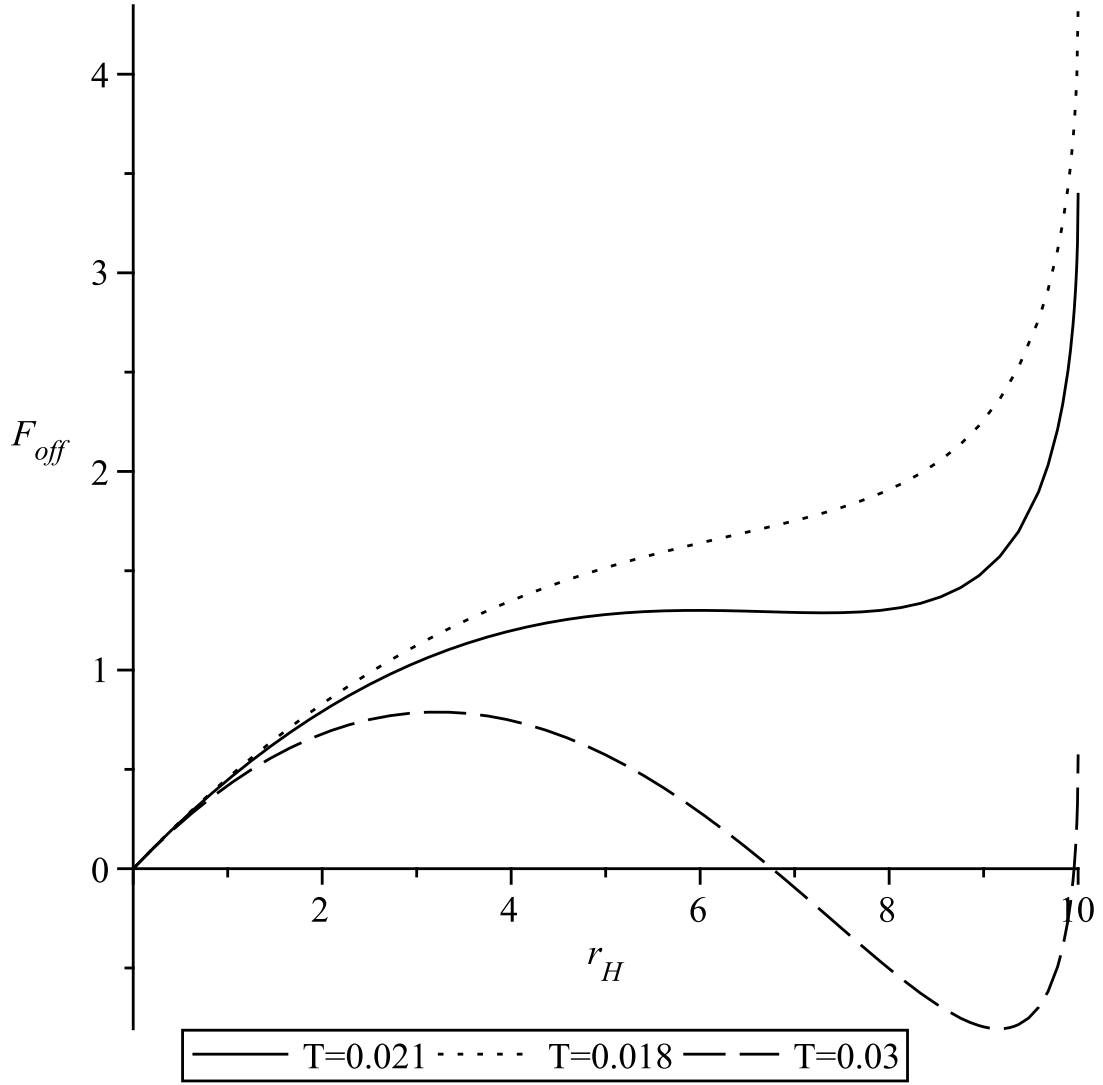}
  \caption{The dotted, solid and dashed curves correspond to the
  dependence of the off-shell free energy of the Schwarzschild black hole
  with a global monopole on the horizon with $8\pi G\eta^{2}\approx10^{-5}$ and $r=10$
  for $T=0.018, 0.021, 0.03$ respectively.}
\end{figure}

\newpage
\begin{figure}
\setlength{\belowcaptionskip}{10pt} \centering
  \includegraphics[width=15cm]{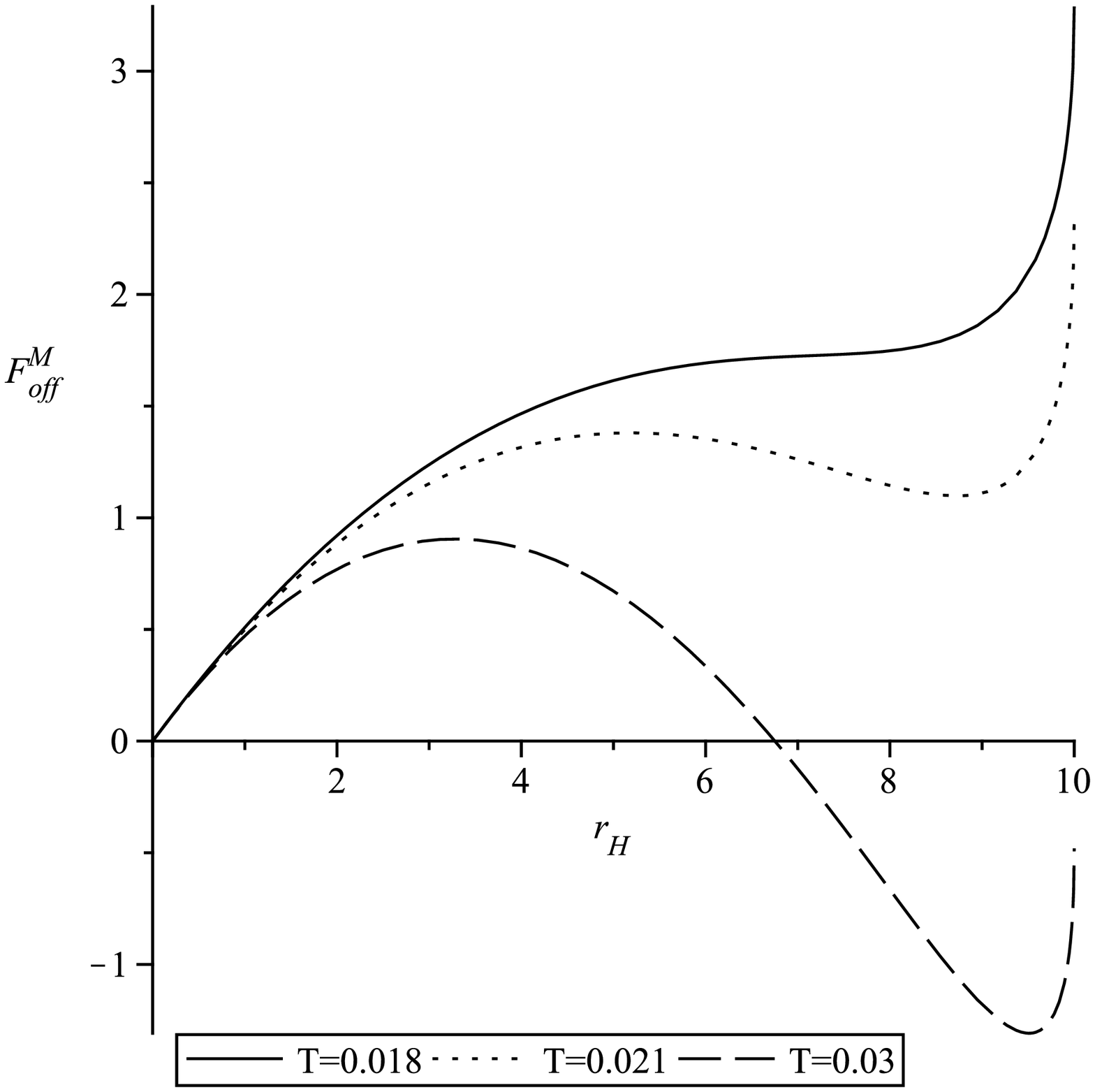}
  \caption{The solid, dot and dashed curves correspond to the
  dependence of the off-shell free energy of the Schwarzschild black hole
  with an $f(R)$ global monopole on the horizon with
  $8\pi G\eta^{2}\approx10^{-5}$, $r=10$ and $\psi_{0}=0.02$
  for $T=0.018, 0.021, 0.03$ respectively.}
\end{figure}


\newpage
\begin{thebibliography}{99}
\bibitem{Bekenstein}J. D. Bekenstein, Lett. Nuovo Cim. 4(1972)737
\bibitem{Bekenstein}J. D. Bekenstein, Phys. Rev. D7(1973)2333
\bibitem{Bekenstein}J. D. Bekenstein, Phys. Rev. D9(1974)3292
\bibitem{Hawking}S. W. Hawking, Commun. Math. Phys. 43(1975)199
\bibitem{Stephens}G. J. Stephens, B. L. Hu, Int. J. Theor. Phys.
40(2001)2183
\bibitem{Kim}W. Kim, E. J. Son, M. Yoon, JHEP 0804(2008)042
\bibitem{Cai}R. G. Cai, L. M. Cao, N. Ohta, JHEP 1004(2010)082
\bibitem{Myung}Y. S. Myung, Y. Kim, Y. Park, Phys. Rev.
D78(2008)084002
\bibitem{Lala}A. Lala, D. Roychowdhury, Phys. Rev. D86(2012)084027
\bibitem{Kim}W. Kim, Y. Kim, arXiv: 1207.5318
\bibitem{Kibble}T. W. B. Kibble, J. Phys. A9(1976)1387
\bibitem{Vilenkin}A. Vilenkin, Phys. Rep. 121(1985)263
\bibitem{Barriola}M. Barriola, A. Vilenkin, Phys. Rev. Lett.
63(1989)341
\bibitem{Cheng}H. Cheng, J. Man, Class. Quantum Grav.
28(2011)015001
\bibitem{Buchdahl}H. A. Buchdahl, Non-linear Lagrangians and
cosmological theory, MNRAS 150(1970)1
\bibitem{Nojiri}S. Nojiri, S. D. Odintsov, Phys. Rev. D68(2003)125312
\bibitem{Carrol}S. M. Carroll, V. Duvvuri, M. Trodden, M. S. Turner, Phys.
Rev. D70(2004)043528
\bibitem{Fay}S. Fay, R. Tavakol, S. Tsujikawa, Phys. Rev. D74(2007)063509
\bibitem{Carames}T. R. P. Carames, E. R. B. de Mello, M. E. X.
Guimaraes, Int. J. Mod. Phys. A (2011)
\bibitem{Carames}T. R. P. Carames, E. R. B. de Mello, M. E. X.
Guimaraes, arXiv: 1111.1856
\bibitem{Man}J. Man, H. Cheng, arXiv: 1205.4857
\bibitem{Tolman}R. C. Tolman, Phys. Rev. 35(1930)904

\end{thebibliography}
\end{document}